# Using Ballistocardiography for Sleep Stage Classification

Jiebei Liu, Peter Morris, Krista Nelson, Mehdi Boukhechba

**Abstract—** A practical way of detecting sleep stages has become more necessary as we begin to learn about the vast effects that sleep has on people's lives. The current methods of sleep stage detection are expensive, invasive to a person's sleep, and not practical in a modern home setting. While the method of detecting sleep stages via the monitoring of brain activity, muscle activity, and eye movement, through electroencephalogram in a lab setting, provide the gold standard for detection, this paper aims to investigate a new method that will allow a person to gain similar insight and results with no obtrusion to their normal sleeping habits. Ballistocardiography (BCG) is a non-invasive sensing technology that collects information by measuring the ballistic forces generated by the heart. Using features extracted from BCG such as time of usage, heart rate, respiration rate, relative stroke volume, and heart rate variability, we propose to implement a sleep stage detection algorithm and compare it against sleep stages extracted from a Fitbit Sense Smart Watch. The accessibility, ease of use, and relatively-low cost of the BCG offers many applications and advantages for using this device. By standardizing this device, people will be able to benefit from the BCG in analyzing their own sleep patterns and draw conclusions on their sleep efficiency. This work demonstrates the feasibility of using BCG for an accurate and non-invasive sleep monitoring method that can be set up in the comfort of a one's personal sleep environment.

―――――――――― ◆ ――――――――――

## 1 INTRODUCTION

Ballistocardiography (BCG) is a mobile sensing technique for quantifying the mechanistic motion of the body (displacement, velocity or acceleration) deriving from the heart's ejection of blood into the aorta and surrounding tissues, respritation, or body motion [1]. The first instrument of its kind was invented in the 1930s. It fell out of favor as a research tool in the 1980s and 90s partly because the extant BCG sensing systems and their associated analytics were non-standardized and more advanced technologies like the more accurate electrocardiography (ECG) gained momentum, becoming incorporated into the polysomnography (PSG) as the gold standard for sleep testing [2]. However, PSG is expensive, cumbersome, and requires expert supervision, making BCG a promising alternative. The modern-day ballistocardiograph typically relies on an ultrasensitive accelerometer but may also consist of a force plate, piezoelectric sensor, or electromechanical film as the source [3].

The waveform of BCG for each heartbeat is characterized by an "IJK" complex at the extremes of signal, which maps out-of-phase onto the QRS complex seen in ECG. Other than the knowledge that the body recoils with each contraction of the heart, the physiological genesis of the normalized amplitudes has long not been wholly definitive. Pinheiro et al. describe the BCG waveform from the perspective of the 1956 Committee on Ballistocardiographic Terminology, a panel of scientists who theorized that the HIJK waves are linked to systole and the smaller LMN waves to diastole, the two phases of a cardiac cycle [4]. Viewing this conclusion to be based on insufficient evidence, Kim et al. argue in a modern fluid mechanics model that the shape originates from the blood pressure gradients in the ascending and descending aorta, the tube-like structure through which oxygenated blood passes from the left ventricle of the heart before being partitioned into peripheral arteries [5]. With regard to its overall accuracy for recognizing heartbeats and beat-to-beat intervals, Mora et al. produced highly precise and sensitive J-peak and J-J interval algorithms without the need for concurrent supervision ECG data [6].

In the last two decades, there has been a resurgence in clinical engineering focus on BCG due to its potential for at-home usage without body contact and the relative ease of installation. An intensely researched application of BCG is toward monitoring sleep from staging of light, deep, and REM, to overall quality in a noninvasive fashion. Sleep has been proven to modulate the autonomic nervous system, with fluctuations between the sympathetic and parasympathetic NS, extending into cardiorespiratory and vascular changes and function that can be systematically quantified and correlated with sleep metrics [7]. In this area of medicine, BCG sensing could serve as a bellwether of such neurological disorders as obstructive sleep apnea and hypopnea, chronic insomnia, hypersomnia (daytime sleepiness), parasomnia (bizarre sleeping behaviors), and nocturnal disturbances. Zink et al. collected B2B signals via a charged polymer BCG foil placed under bed sheets, superimposing results with breathing activity in patients suspected of sleep apnea. Their multidimensional algorithm on BCG data had as one of its components a quality index (QI) that excluded data for two consecutive waves that are discrepant enough to be judged artifactual. They showed good agreement with the output of ECG [8]. Cmir and Studnička directed convolutional neural networks toward BCG-detected time delay between R-wave and pulse in order to study abnormal vital signs that can be associated with breathing disorders [9]. Mitsukura et al. explored BCG-detected sleep stages through mathematical programming, placing aluminum alloy sensors under each bed leg [10].

The presence of motion-induced artifacts found in ballistocardiograms represents a common weakness, lowering the signal-to-noise ratio and thereby challenging the diffusion of certified biomedical devices into cardiology and neurology practices [11]. Jaworski et al. highlight the importance of a filtering method to remove these distortions of BCG while retaining a continuous stream of cardiac micromovements [12].

Although BCG has existed for nearly a century, it is a relatively novel approach for sleep management. It is backed in some way by PSG, implying that it has great accuracy in measuring sleep-related problems [13]. All topics related to BCG's sleep classification, sleep-related symptom monitoring, and BCG signal processing show very good prospects. The main objective for the work presented here is to monitor sleep through external non-invasive sensors and to evaluate the sleeping efficiency using a non-training algorithm and classify different sleeping stages using machine-learning algorithms.

## 2   BACKGROUND

Humans spend about a third of their lifespan asleep. As per the American Association of Sleep Medicine, this nocturnal neural activity possesses a cyclic structure with an average period of 90 minutes that can be broken into three main groups: rapid eye movement (REM), non-rapid eye movement (NREM), and intermediate wakefulness. NREM is further subdivided into stages 1, 2, and 3 (formerly 3 and 4), each with its own polysomnogram character. N1 and N2 are considered light sleep, together lasting from 30 to 70 minutes per cycle. N3 is deep, delta wave, or slow wave sleep (SWS), about 20-40 minutes. REM is when dreaming occurs and its duration increases over the night (from 10 minutes to sometimes over an hour) [14]. It is regarded as paradoxical sleep because of the brain waves' similarity to a wakeful state [15]. One typically enters from arousal to NREM, when HR decreases from N1 to N3, before reaching REM when HR is more erratic [16].

Sleep quality is a rather complex construct of restorativeness, and it would be reductive to score it without having a thorough grasp of the validity of all the parameters that account for it. Historically, sleep quality has been measured via subjective techniques such as the Consensus Sleep Diary (CSD) or the Pittsburgh Sleep Quality Index (PSQI) that are often incorporated into behavioral treatments of disordered rest. More on the objective side are polysomnography and actigraphy, the latter of which acquires sleep-wake cycles and circadian rhythm patterns [17]. The CSD and the PSQI both use standardized measures of sleep quality in questionnaires to divide its subjects into 'good' or 'poor' sleepers whose components are as follows: sleep quality, sleep latency, sleep duration, habitual sleep efficiency, sleep disturbance, use of sleeping medication, and daytime dysfunction. While PSQI has been labeled the optimal self-report of sleep quality, and PSG the gold standard for objective measure of sleep, there is not a strong correlation between PSQI and PSG data results. This is likely because PSQI blends data from an entire month's worth of sleep to determine sleep quality, while the scope of PSG is confined to night-to-night and it must take place in a lab, making multi-night recordings more impractical [18].

Heart rate variability can also provide information about sleep efficiency, and can be used to identify sleep disorders such as sleep apnea and insomnia. Greater HRV during resting wakefulness is associated with better sleep efficiency, assessed by comparing results from PSQI and PSG over the same timeframe. HRV has been determined an independent marker of sleep efficiency during a short resting period, and could proxy for sleep quality, as is attempted in this study [20]. The operational definition of sleep efficiency is inconsistent or misapplied in the literature, and this subjectivity could hinder progress in sleep research. According to Reed and Sacco, it should be calculated as the total time asleep over the duration of sleep episode [19]. In addition, sleep onset latency (SOL) and wake after sleep onset (WASO) have become popular, less controversial indices of sleep performance, which are respectively the time it takes for one to fall asleep after getting in bed, and the time awake estimated from initial sleep onset to getting out of bed [20]. The more severe SOL and WASO are, the more likely one is to meet the criteria for insomnia and other sleep disturbances [21].

In the sections that follow, we will describe our methodology for a sleep study leveraging BCG sensing technology, including a consumer-grade Fitbit as our ground truth. The results are reported for sleep and wake detection according to a previously ballistocardiographic-defined threshold as well as sleep stage estimation computed from machine-learning scripts, and their implications discussed. We will transition to limitations of our work and guidance for future research into the improvement of our scientific design and knowledge of this powerful sensor.

## 3 METHODS

After a thorough search into the market of available BCG devices that have been cited in the literature, the SCA11H by MuRata was chosen as it is promoted for monitoring sleep in research and commercial settings and has an acceptable level of granularity for our purposes. A microelectro-mechanical system (MEMS) sensor, SCA11H generates values of heart rate (HR), heart rate variability (HRV), respiratory rate (RR), and beat-to-beat or R-R peak interval (B2B) at a frequency of 1 Hz through the proprietary BCG mode in the microcontroller. This information was recorded nightly on Python's IDLE from the sample code that Murata makes available, which was configured over home Wifi networks. The device was positioned and calibrated about 8-10 inches away and in line with the chest, parallel to the supine body, and either on top of the bedsheets, under a mattress pad, or fixed to the side of the bed frame. This flexibility in contactless placement demonstrates its utility in sleep monitoring.

To compare against ground truth, photoplethysmogram (PPG) sensing from a Fitbit Sense Smart Watch provided real-time heart rate and motion data from which sleep stages could be automatically inferred and downloaded via the API. The test subjects calibrated the watch to their individual gender, height, weight, and age, and wore the watch on their non-dominant wrist in unison with the data collection from the BCG. Fitbit estimates sleep stages through tracking its users movement and heart rate variability patterns, and inputting these into an algorithm that combines these and infers whether the user is in light sleep, deep sleep, or REM. This validated model served as a baseline for extrapolating sleep stages with the information gathered from the BCG [22].

### 3.1 DATA SUMMARY

The sleeping data was collected over nine nights with two people (1 male and 1 female) aged around 24. For all 8 nights, the SCA11H sensor output signals (BCG) and the Fitbit sleep state reference value (PPG) were recorded at 1-second intervals. This information was utilized to conduct further sleep and wake classification, data analysis, and investigate the efficacy of a variety of predictive models. The average sleep length of these 8 days is 29,800 seconds (8.28 hours). The average of each features per day is in Fig.1

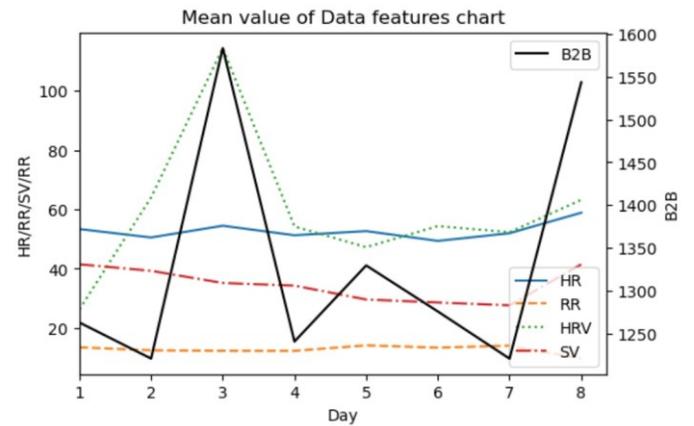

Figure 1: The mean value of each features

### 3.2 ASLEEP-AWAKE ALGORITHM

The analytical techniques in this paper for the BCG output are Inspired by the work of Park et al. [23] and Jung et al. [24]. Each night of data was separated into 30-second epochs that were designated either "awake" or "asleep." This was accomplished through a training-free binary segmentation algorithm in which a moving threshold from the previous 3 minutes of heart rate (beats per minutes) dictated these states for each epoch. In particular, the threshold was calculated as the mean of the previous 3 minutes of HR plus a scalar times their standard deviation. The first 3 minutes of logged data were tagged awake. When HR samples were taken from the first 3 minutes of data (i.e., minutes 3 to 5:30), the scalar was –1; thereafter it was equal to 2 to be within a 95% confidence interval. If more than half of the samples (1 Hz) per epoch were below the threshold, it was designated asleep except in the event of more than ten zeros owing to corruptive body movement.

Sometimes SCA11H could not reliably provide HR data, in which it would either insert a zero (regardless of the other nonzero parameters) or entirely omit those seconds of runtime (all of the parameters gone), the latter of which meant the connection was lost momentarily. As Jung asserts, a zero HR reading indicates the BCG waveform was defective in response to motion. If more than ten samples of HR per 30-second epoch were zero, then it was interpreted as awake. The sleep efficiency could then be computed jointly from body movement and threshold classifications, or the ratio of the number of epochs in the asleep state to the total number of bedtime epochs.

### 3.3 ALL STAGES CLASSIFICATION ALGORITHM

Another way is training various classification models to find a high-accuracy model that can differentiate distinct sleep stages using data from the Murata BCG sensor's BCG mode. Before training, any missing values had to pass imputation, which means they will be filled by their previous or next value. This method can be employed because the body signal during sleep will not differ significantly over one second. Then, based on the findings of the PCA analysis (Fig.1), five features including HR, RR, SV, HRV, and B2B, are set aside for future investigation.

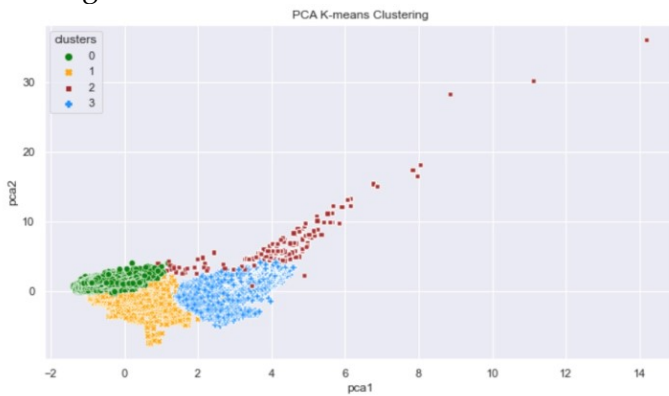

Fig. 2: PCA Analysis (the explained variance ratio of five features are 0.35, 0.25, 0.17, 0.13, 0.1)

The time-series data were then separated into epochs using 10-second sliding windows. Each epoch was given a label based on Fitbit's sleeping stage classification. When the 10s epoch spans two stages, it was discarded because these epochs can completely represent neither sleep stage. The features used in the model training include the mean, median, maximum, minimum, standard deviation, and 75% percentile of HR, RR, SV, B2B, and HRV in each 10s-epoch. There are 25,282 wake stage epochs, 46,980 Rem stage, 120,742 light stage and 43,580 deep sleep stage epochs.

The data are divided into 80% for training and 20% for testing, with 5-folds cross-validation. The classification algorithms include: Decision Tree, k-nearest neighbors (KNN), Random Forest, Naive Bayes, Gradient Boosting, and Support Vector Machines (SVM). The 6 algorithms were implemented using sklearn package in python. Results presented in Table 1 suggest that Random Forest stood out as the best model with an average F1 score of 0.987 followed by decision trees then KNN.

|  | RMSE | Accuracy | F1-Score |
|---|---|---|---|
| Decision Tree | 0.26 | 0.856 | 0.839 |
| KNN | 0.478 | 0.749 | 0.718 |
| **Random Forest** | **0.31** | **0.942** | **0.937** |
| NaiveBayes | 1.01 | 0.302 | 0.25 |
| Gradient Boosting | 0.713 | 0.649 | 0.602 |
| SVM | 0.891 | 0.521 | 0.333 |

Table 1: Performance for different classification algorithms.

### 4 RESULTS & DISCUSSION

The first goal of this study was to measure sleep efficiency as a function of the number of epochs asleep over the total epochs recorded per night from the moment of getting in bed. An example of this time series can be visualized in a 30-minute period of confirmed sleep in Fig 3.

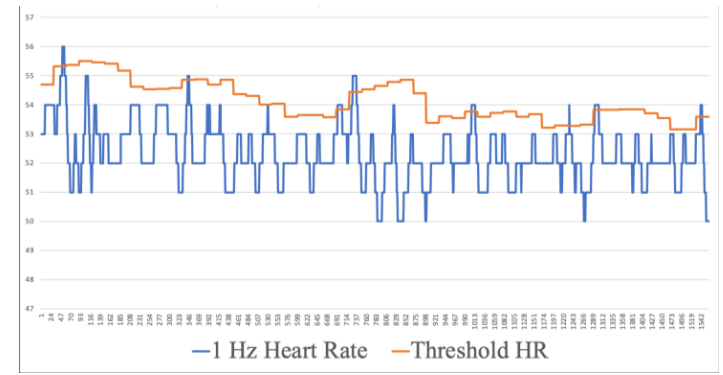

Fig. 3. Heart Rate (Beats per Minute) Oscillations (1 Hz) compared to Sleep/Wake Threshold (0.0333 Hz) from 4am to 4:30am.

Even though some heart rates spiked above the threshold, they were not sufficiently sustained over the 30-second subintervals to be called "awake." Thus within these thirty minutes the subject was most probably in one or more unconscious stages.

Pearson's $r$ for this parameter is 0.897 with $p$-value 0.0025. Although there were 14 attempts at collecting simultaneous BCG and PPG data, only eight nights were eligible for the simple threshold detection approach. As seen in Fig. 4, the BCG sensor achieves reasonable concordance with the Fitbit PPG in in the mean and range sleep efficiency.

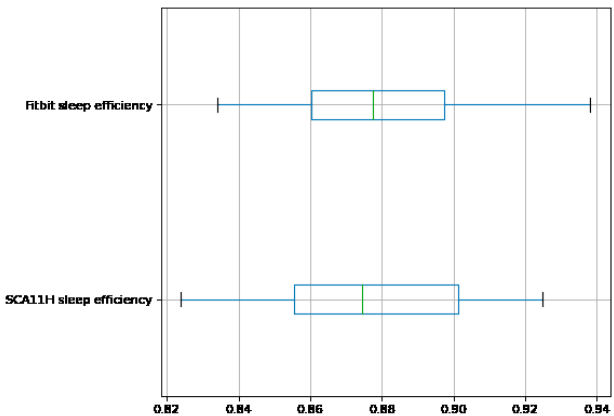

Fig. 4: Box plot of sleep efficiency of BCG vs. Fitbit.

The second goal of this study was to obtain the best classification algorithm. Based on this, the confusion matrix is obtained with the Fitbit PPG(column) and the machine learning predicted values of the sleep phases obtained with BCG(row). Since the highest value between columns and rows are in the desired positions for a great performance. It is concluded that the result of RandomForest presents the best performance. The confusion matrix of Random Forest is in Figure 5.

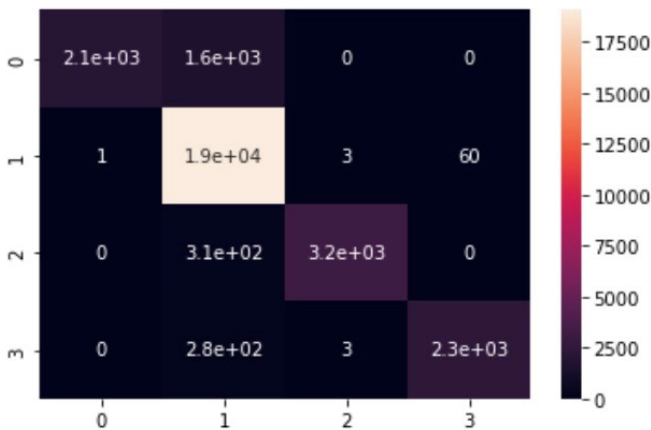

Fig.5: Confusion matrix of predicting sleep stages using Random Forest.

We have also obtained figures to compare predictions with the labeled data. In particular, in Figure 6, we compare the sleep phase indicated by Fitbit with the predicted one using the random forest classifier. When analyzing this figure, it is noticeable that the prediction sleeping stage is similar to the fitbit generated one, as expected. Thus, the random forest classifier can be used to do further analysis.

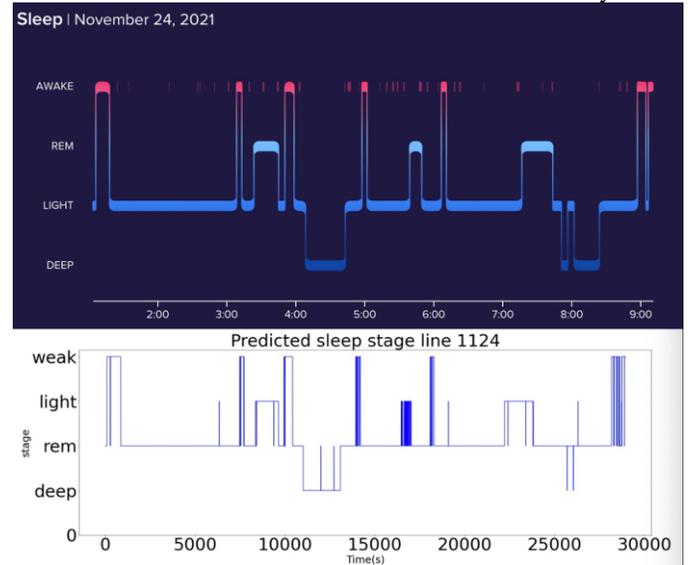

Fig.6: The comparison between Fitbit hypnogram(upper) and the BCG-predicted hypnogram(bottom).

It should be noted that these prediction algorithms are only effective for non-medical applications, and the main goal is to help people realize and understand each sleep phase and try to improve their sleep quality.

## 5 LIMITATIONS

**BCG Mode**: The estimation of sleep stage solely from post-processed BCG signals of HR, HRV, B2B, and/or RR could not be accomplished because all the works we found investigated the parameters that constitute the raw signal of, e.g., HF:LF ratio. Since we were only in BCG mode, we could not exert a validated algorithm on the empirical results.

**Internet Connection**: Loss of Wifi and/or SCA11H connection during sleep commonly led to missing, irretrievable data. A few nights had 10,000 or more seconds skipped in the Python IDE for no apparent reason and unfortunately had to be discarded even though good Fitbit data were obtained. While the users were sleeping, inevitably no intervention could be taken to restart the system. Difficulty also occurred in the Wifi connection of the BCG device to the computers being used for data collection. This resulted in a number of nights in which data collection was either interrupted partway through the night or was not able to be collected at all. Because of these complications, and the limited timespan of this project, the amount of data collected was not as large as what would have been preferred.

**Data Size**: Only two young people participated in our sleep study. And only 8 days' data are complete and accurate enough to be used in both methods. As a result, the model we train can be overly customized because it only learns different sleep patterns from two participants rather than many people of all ages. As the amount of data in subsequent studies grows, this problem can be addressed.

**Unbalanced data**. The data from different sleep stages is uneven, which could lead to some stages being overfitted. The results also show that random forest's performance in identifying light sleep stages is clearly inferior to that of other stages. This situation should be improved by creating synthetic samples that sample attributes from instances in the minority class at random.

**Nature of Fitbit**. First, we discovered that the sleep stage data does not contain the wake-up data of less than 3 minutes between the first sleep and the entire wake-up when using Fitbit. This will significantly impact how we compare sleep/wake classification algorithms using figures. Similarly, this circumstance will also impact the categorization machine learning model's accuracy. Second, when monitoring sleep, Fitbit utilizes its own algorithm to record the time of falling asleep and waking up automatically. However, on some specific dates, sleep data is cut off during light or deep sleep stages, which reduces the accuracy of total sleep time and sleep efficiency calculations. As a result, Fitbit's data cannot be considered an entirely accurate measurement. Other studies that use PSG as an indication may be more consistent and accurate.

## 6 Future Investigation

There are many future applications that could arise from the research presented in this study. The integration to electronic devices from the BCG Murata SCA11H, as allowed through its PCBA-module, presents the ability for straightforward connection to mobile devices. At-home BCG data collection for the common consumer would allow for instantaneous access to information on the user's heart rate, heart rate variability, respiratory rate, and beat-to-beat or R-R peak intervals. Having all of these capabilities at hand, it could be possible to integrate the BCG technology into a telehealth or mobile health application for the storing and analysis of this collected data. In doing so, it would become possible to instantly share information regarding a person's sleeping habits, or simply status of bed occupation, with a doctor, caregiver, or researcher.

This application could be beneficial in an at-home setting, a care home setting, or a hospital setting. An application of this kind would provide the user with instant feedback on their sleep stages and sleep efficiency through implemented algorithms. BCG sensing would also provide opportunity for health analysis that could be deployed to detect a variety of sleeping-related health issues; namely, with appropriate interpretation, obstructive sleep apnea and hypopnea, chronic insomnia, hypersomnia, parasomnia, and nocturnal disturbances, as well as more tangential health complications such as stress and cardiovascular diseases.

A technology of this sort could also assist other research areas such as trying to understand how often, and at what times, a person with dementia leaves their bed during the night, for example. There are many other neurological and psychological areas of research that could utilize the information made available by this device. This research has shown that the integration of BCG with a user-friendly interface would allow for a multitude of applications for personal use, for medical advancements, and for contributions to other research areas.

## 7 Conclusion

As demonstrated in this study, the field of ballistocardiographic sensing for sleep stage classification shows early promise, despite challenges faced during the study. This non-invasive sleep monitoring device has the capability to be used as a determinant of sleep stages and sleep efficiency through its collection of timestamps, heart rate, respiration rate, relative stroke volume, and heart rate variability, when combined with the appropriate algorithm. When compared to the gold standard, polysomnography as used for sleep testing, it is clear that this method is less expensive, less cumbersome, and requires less supervision during experimentation. However, the results of this study remind us that we are still a long way from entirely replacing the outputs of PSG recordings. While there is the clear benefit of being able to practice BCG in an at-home setting, as opposed to the laboratory setting that is required by the PSG, it is also clear that the limitations in what is able to be monitored by the BCG cause this device to result in less accuracy than the PSG in determining sleep stages.

Overall, this study shows an inexpensive and minimally invasive method of collecting data that provides many benefits to its users in terms of understanding their health and overall wellbeing. The use of data that is patient-specific, accurate, and minimally burdensome provides for future healthcare models that are smarter and more personalized than previously available to the average person. There, however, remains some obstacles as it pertains to interpretation of data that still needs to be understood and worked through. The training of data and machine learning predicted values of sleep phases with the BCG could be improved with more data and more analysis. Addressing this issue would require more time for this study, and a greater understanding of machine learning techniques. Despite the limitations faced, this study does provide a strong foundation for understanding sleep staging with BCG, and provides for many further areas of research that could benefit different groups of people, as well as lead to future research and advancements in telehealth, mHealth, and health care technologies.